\documentclass[prd,preprint,
superscriptaddress,showpacs,nofootinbib,%
tightenlines
]{revtex4}
\usepackage{epsfig}
\usepackage{color}
\usepackage{amssymb}

\newcommand{\ben}{\begin{displaymath}}
\newcommand{\een}{\end{displaymath}}
\newcommand{\be}{\begin{equation}}
\newcommand{\ee}{\end{equation}}
\newcommand{\bea}{\begin{eqnarray}}
\newcommand{\eea}{\end{eqnarray}}
\begin{document}
\title{Once more on the Higgs decay into two photons}
\author{J.~Gegelia}
 \affiliation{Institute for Advanced Simulation, Institut f\"ur Kernphysik
   and J\"ulich Center for Hadron Physics, Forschungszentrum J\"ulich, D-52425 J\"ulich,
Germany}
\affiliation{Tbilisi State  University,  0186 Tbilisi,
 Georgia}
 \author{Ulf-G.~Mei\ss ner}
 \affiliation{Helmholtz Institut f\"ur Strahlen- und Kernphysik and Bethe
   Center for Theoretical Physics, Universit\"at Bonn, D-53115 Bonn, Germany}
 \affiliation{Institute for Advanced Simulation, Institut f\"ur Kernphysik
   and J\"ulich Center for Hadron Physics, Forschungszentrum J\"ulich, D-52425 J\"ulich,
Germany}
 \date{April 13, 2018}
\begin{abstract}

We comment on the recently reiterated claim that the contribution of the W-boson loop to the 
Higgs boson decay into two photons leads to different expressions in the $R_\xi$ gauge and the 
unitary gauge. By applying a gauge-symmetry preserving regularization with higher-order 
covariant derivatives we reproduce once again the ``classical'' gauge-independent result.

\end{abstract}




\maketitle

The original calculations of the W-boson loop contribution to the Higgs boson decay into two 
photons \cite{Ellis:1975ap,Ioffe:1976sd,Shifman:1979eb}
have been challenged in Refs.~\cite{Gastmans:2011ks,Gastmans:2011wh,Gastmans:2015vyh}. 
In Ref.~\cite{Christova:2014mea} it has been argued that 
the dispersion theory calculation confirms the discrepancy. However, the careful and 
detailed studies of Ref.~\cite{Melnikov:2016nvo} revealed that 
unregulated and unsubtracted results in the unitary gauge are incorrect in spite 
of being finite. 

The issue of the gauge (in)dependence of the Higgs decay amplitude has been raised again 
in a recent publication~\cite{Wu:2017rxt} where it has been claimed that the results of the 
$R_\xi$ gauge and the unitary gauge are explicitly verified to be different.

\begin{figure}[htb]
\epsfig{file=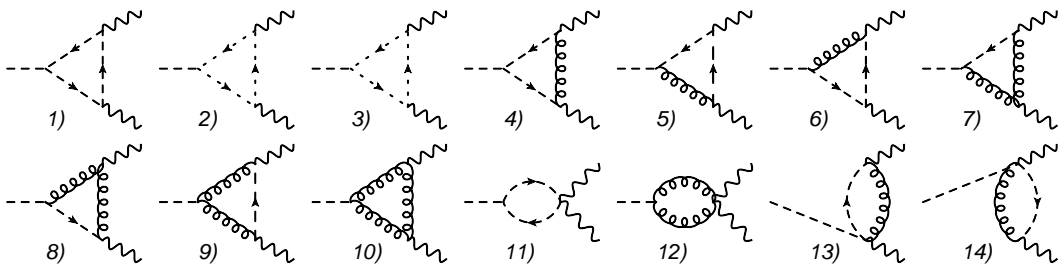, width=14truecm}
\caption[]{\label{Hgg:fig} One-loop diagrams of the W-boson loop contribution in the Higgs boson 
decay into two photons. Crossed diagrams are not shown. Curved, wiggled, dashed, dashed with arrows 
and dotted lines correspond to the photon, W-boson, Higgs scalar, Goldstone bosons and the 
Faddeev-Popov ghosts, respectively. Notice that diagrams 2) and 3) correspond to two 
different ghost lines (see Ref.~\cite{Aoki:1982ed} for details).}
\end{figure}

Using the Feynman rules (and notations) of Ref.~\cite{Aoki:1982ed} we obtained that  
in $R_\xi$ gauge all ultraviolet divergences of one-loop diagrams appearing in  the W-boson loop 
contribution to the Higgs boson decay into two photons (diagrams are shown in Fig.~\ref{Hgg:fig}) 
cancel at the level of integrands except  
\begin{equation}
\frac{e^3 M_Z  \left(2 (D-1) M_W^2+m_{\phi
   }^2\right)}{M_W
   \sqrt{M_Z^2-M_W^2}} \int \frac{d^Dq}{(2 \pi)^D}
   \frac{4 q^{\mu } q^{\nu }-q^2
   g^{\mu \nu }}{[q^2-M_W^2]^3 },
\label{decayampl}
\end{equation}
with $M_Z, M_W$ and $m_\phi$ the masses of the Z-boson, the W-boson and the Higgs
particle, respectively, and $e$ is the conventional  electromagnetic coupling constant.
This integral is also finite, however, only after  the loop integration has been carried out.
Note that  while we explicitely worked in $D$ dimensions, one can also do the algebra in 
four space-time dimensions which amounts to setting $D=4$.
This is exactly the same integral which has been identified as the source of the discrepancy 
between the unitary  and $R_\xi$ gauges \cite{Gastmans:2011ks,Gastmans:2011wh,Gastmans:2015vyh} 
(see also the careful derivation of Ref.~\cite{Li:2017hnv}).
The problem with the finite loop integral in Eq.~(\ref{decayampl}) is that it is a difference 
of two logarithmically divergent integrals and cannot be calculated 
without regularization (see, e.g., Ref.~\cite{Weinzierl:2014iaa}). While the divergent parts 
of these two integrals have the same 
coefficient for any Lorentz-invariant regularization, different regularizations 
generate different finite pieces and therefore the final result depends on the applied 
regularization scheme.  
Thus the problem actually is not with the unitary  and $R_\xi$ gauges leading to different results, 
but rather the result being dependent on the way we calculate the divergent integrals. 
If we deal with the expression of Eq.~(\ref{decayampl}) the same way as done in 
Refs.~\cite{Gastmans:2011ks,Gastmans:2011wh,Gastmans:2015vyh} we get a result different 
from that of the dimensional regularization also in $R_\xi$ gauge.

To verify once again 
that the dimensional regularization leads to a correct result and that the problem is caused 
by the incorrect treatment of the integral of Eq.~(\ref{decayampl}),
we applied a gauge symmetry preserving regularization with higher-order covariant 
derivatives \cite{Faddeev:1980be} to the electroweak theory by adding the following 
regularizing terms to the Lagrangian (we use the notations and the parametrization of 
Ref.~\cite{Aoki:1982ed})
\begin{eqnarray} 
{\cal L}_{HD} & = & g_{HD}\,D^{ab}_{\mu}F^b_{\nu\lambda}D^{ac,\mu} F^{c\nu\lambda}\,,\nonumber\\
D^{a b}_{\mu} & = & \delta^{a b} \partial_\mu -g f^{abc} W^c_\mu\,,
\label{HDT}
\end{eqnarray}
where $W^a_\mu$ is the triplet of SU(2) vector bosons and $F^a_{\mu\nu}=
\partial_\mu W^a_\nu-\partial_\nu W^a_\mu + gf^{abc} W_\mu^b W_\nu^c$ is the corresponding
field strength tensor. 
The addition of the term of Eq.~(\ref{HDT}) to the Lagrangian of the 
electroweak theory leads to modifications of the Feynman rules. Below we specify only those 
which are relevant for our calculation. The modified propagator of  the W-boson has the form
\begin{equation}
\frac{1}{4 g_{\text{HD}} (k^2)^2 + k^2 + i\,\epsilon  -M_W^2} \left[
g^{\mu \nu} 
-\frac{k^{\mu } k^{\nu }
   \left( 1 - \alpha (1 +4 g_{\text{HD}} k^2
   ) \right)}{ k^2+i\,\epsilon -\alpha  M_W^2 } \right]~,
\label{WPr}
\end{equation}
with $\alpha$ the gauge parameter associated with the W-boson (for more 
details, we refer again to Ref.~\cite{Aoki:1982ed}).
There is an additional $A_\alpha(k)  W^-_\beta(p) W^+_\gamma(q) $ vertex with all momenta incoming,
\begin{eqnarray}
&& -4 e g_{\text{HD}} \, g^{\alpha \beta } \left[ p^{\gamma } (2
   \text{\textit{k}}\cdot p+p\cdot
   q)-\text{\textit{k}}^{\gamma } (2
   \text{\textit{k}}\cdot p+\text{\textit{k}}\cdot
   q)\right] 
   \nonumber\\
   &&
   \ \ +4 e g_{\text{HD}} \, g^{\beta \gamma }
   \left[ q^{\alpha } (\text{\textit{k}}\cdot q+2 p\cdot
   q)-p^{\alpha } (\text{\textit{k}}\cdot p+2 p\cdot
   q)\right]
      \nonumber\\
   &&
   \ \  +4 e g_{\text{HD}} \, g^{\alpha \gamma }
   \left[  \text{\textit{k}}^{\beta }
   (\text{\textit{k}}\cdot p+2 \text{\textit{k}}\cdot
   q)-q^{\beta } (2 \text{\textit{k}}\cdot q+p\cdot
   q)\right)
      \nonumber\\
   &&
   \ \  +4 e g_{\text{HD}} \, 
   \left[\text{\textit{k}}^{\beta }
   \left( \text{\textit{k}}^{\gamma } \left(p^{\alpha
   }-q^{\alpha }\right)-p^{\alpha } p^{\gamma
   }\right)+q^{\beta } \left(q^{\alpha }
   \left(\text{\textit{k}}^{\gamma }-p^{\gamma
   }\right)+p^{\alpha } p^{\gamma }\right)\right],
 \label{AWW}
\end{eqnarray}
and an additional $W^+_\alpha(p)  W^-_\beta(q)  A_\gamma(r) A_\delta(k)  $ vertex with all momenta incoming,
\begin{eqnarray}
&& 4 e^2 g_{\text{HD}} \biggl[ g^{\alpha \gamma } g^{\beta \delta }
   (2 \text{\textit{k}}\cdot q+\text{\textit{k}}\cdot
   r+p\cdot q+2 p\cdot r)  +
   g^{\alpha \delta } g^{\beta \gamma } (2
   \text{\textit{k}}\cdot p+\text{\textit{k}}\cdot
   r+p\cdot q+2 q\cdot r)
   \nonumber\\
   &&
 \ \    -  g^{\alpha
   \beta } g^{\gamma \delta } (\text{\textit{k}}\cdot
   p+\text{\textit{k}}\cdot q+4 \text{\textit{k}}\cdot r+4
   p\cdot q+p\cdot r+q\cdot r)\biggr]
   \nonumber\\
&&
 \ \  -4 e^2 g_{\text{HD}} \biggl[ -2 g^{\gamma \delta }
   \text{\textit{k}}^{\alpha } \text{\textit{k}}^{\beta
   }+p^{\delta } g^{\beta \gamma }
   \text{\textit{k}}^{\alpha }+p^{\beta } g^{\gamma \delta
   } \text{\textit{k}}^{\alpha }+\text{\textit{k}}^{\gamma
   } \left(
   g^{\alpha \delta } \text{\textit{k}}^{\beta
   }+g^{\beta \delta } \text{\textit{k}}^{\alpha
   }+p^{\delta } g^{\alpha \beta } 
   \right.
   \nonumber\\
 &&  \left.
 \ \    -2 p^{\beta } g^{\alpha
   \delta }+q^{\delta } g^{\alpha \beta }-2 q^{\alpha }
   g^{\beta \delta }-2 r^{\delta } g^{\alpha \beta
   }-r^{\beta } g^{\alpha \delta }-r^{\alpha } g^{\beta
   \delta }\right)
   +q^{\delta } g^{\alpha \gamma }
   \text{\textit{k}}^{\beta }+q^{\alpha } g^{\gamma \delta
   } \text{\textit{k}}^{\beta }
   \nonumber\\
 &&  
 \ \    -r^{\delta } g^{\alpha
   \gamma } \text{\textit{k}}^{\beta }-r^{\delta }
   g^{\beta \gamma } \text{\textit{k}}^{\alpha }+2
   r^{\beta } g^{\gamma \delta } \text{\textit{k}}^{\alpha
   }+2 r^{\alpha } g^{\gamma \delta }
   \text{\textit{k}}^{\beta }+2 p^{\delta } q^{\gamma }
   g^{\alpha \beta }-p^{\delta } q^{\alpha } g^{\beta
   \gamma }-p^{\beta } q^{\delta } g^{\alpha \gamma
   }
   \nonumber\\
 &&  
 \ \    -p^{\beta } q^{\gamma } g^{\alpha \delta }-2 p^{\beta
   } q^{\alpha } g^{\gamma \delta }-2 p^{\beta } r^{\delta
   } g^{\alpha \gamma }+p^{\beta } r^{\alpha } g^{\gamma
   \delta }+p^{\beta } p^{\delta } g^{\alpha \gamma
   }
   \nonumber\\
 &&  
  \ \   +p^{\gamma } \left(-2 p^{\delta } g^{\alpha \beta
   }+p^{\beta } g^{\alpha \delta }+2 q^{\delta } g^{\alpha
   \beta }-q^{\alpha } g^{\beta \delta }+r^{\delta }
   g^{\alpha \beta }+r^{\alpha } g^{\beta \delta
   }\right)+q^{\gamma } r^{\delta } g^{\alpha \beta
   }+q^{\gamma } r^{\beta } g^{\alpha \delta }
   \nonumber\\
 &&  
 \ \    -2 q^{\alpha
   } r^{\delta } g^{\beta \gamma }+q^{\alpha } r^{\beta }
   g^{\gamma \delta }-2 q^{\gamma } q^{\delta } g^{\alpha
   \beta }+q^{\alpha } q^{\delta } g^{\beta \gamma
   }+q^{\alpha } q^{\gamma } g^{\beta \delta }+r^{\beta }
   r^{\delta } g^{\alpha \gamma }
   \nonumber\\
 &&  
 \ \    +r^{\alpha } r^{\delta }
   g^{\beta \gamma }-2 r^{\alpha } r^{\beta } g^{\gamma
   \delta }\biggr].    
    \label{AAWW}
\end{eqnarray}
We add these two vertices to the corresponding expressions of the Feynman rules specified 
in Ref.~\cite{Aoki:1982ed} so that the topologies and the number of Feynman diagrams remain the same.
All other additional vertices generated by the term of  Eq.~(\ref{HDT}) are not relevant 
for the current calculation.
There are twenty six one-loop diagrams in the W-loop contribution to the Higgs boson 
decay  into two photons, shown in Fig.~\ref{Hgg:fig}. For $\alpha = 0$ all diagrams containing 
at least one W-boson propagator are finite for non-vanishing $g_{HD}$.
Diagrams 1), 2), 3), their crossed partners and diagram  11) are regularized by subtracting 
the analogous loop diagrams with propagators with a heavy mass $\Lambda$, amounting to 
gauge symmetry  preserving Pauli-Villars regularization \cite{Faddeev:1980be}. For convenience 
in the calculations we take $g_{HD}=1/(4 \Lambda^2)$ so that the removed regulator limit is 
obtained by taking the limit $\Lambda\to \infty$ after 
performing the loop integration (and subtracting divergences - if there were any).

In the calculation of the loop diagrams we apply the method of dimentional counting 
of Ref.~\cite{Gegelia:1994zz}, that is similar to the ``strategy of regions'' of 
Refs.~\cite{Beneke:1997zp,Smirnov:2002pj}. This method allows to represent each regulated loop diagram 
in four space-time dimensions as the sum of 
two expressions, both calculated by applying dimensional regularization. The first expression 
for each diagram is obtained by expanding the integrand of the one-loop integral in 
inverse powers of $\Lambda$ and interchanging the integration and summation. For 
$\Lambda\to \infty$ these expressions exactly coincide to the standard Feynman diagrams 
obtained using the Feynman rules of Ref.~\cite{Aoki:1982ed} and applying  dimensional regularization. 
The corresponding second part for each diagram 
is obtained by rescaling the integration variable $k\to q \Lambda$, expanding the 
resulting integrand in inverse powers of $\Lambda$ and interchanging the integration and the 
summation. 

Let us briefly demonstrate the method of dimensional counting for a simple massless 
one-loop integral regulated using a Pauli-Villars type regulator,
\begin{eqnarray}
I & = & \int \frac{d^4 k }{(2 \pi)^4}  
  \frac{-\Lambda ^2}{[k^2-\Lambda^2+i\epsilon]}\frac{1}{[k^2+i\epsilon] [(k+p)^2+i\epsilon]} \Rightarrow I_1+I_2\,,\nonumber \\
I_1 &=& \int \frac{d^D k }{(2 \pi)^D}  
 \frac{1}{[k^2+i\epsilon][(k+p)^2+i\epsilon]}+ \frac{1}{\Lambda^2} \int \frac{d^4 k }{(2 \pi)^4}  
  \frac{1}{(k+p)^2+i\epsilon}+{\cal O}\left( \frac{1}{\Lambda^4}\right), \nonumber\\  
  I_2 &=& \Lambda^{D-4}\left( \int \frac{d^D q}{(2 \pi)^D}  
 \frac{-1}{(q^4+i\epsilon)(q^2-1+i\epsilon)} + \frac{1}{\Lambda}\int \frac{d^D q}{(2 \pi)^D}  
 \frac{2 p\cdot q}{(q^6+i\epsilon)(q^2-1+i\epsilon)} +{\cal O}\left( \frac{1}{\Lambda^2}\right) \right) . \nonumber
\label{demInt}
\end{eqnarray}
Calculating the dimensionally regulated integrals $I_1$ and $I_2$ and expanding at $D=4$,
one finds that the $1/(D-4)$ poles cancel and obtains
\begin{equation}
I =  -\frac{i}{16 \pi ^2} \left(-1+\ln \frac{-p^2-i \epsilon}{\Lambda
   ^2}\right) +{\cal O}\left( \frac{1}{\Lambda}\right) .
\label{resI}
\end{equation}

\medskip

Let us return to our original problem.
If the sum of the second parts of all twenty six diagrams is 
non-vanishing in the limit $\Lambda\to \infty$ for $D=4$ that would mean that 
dimensional regularization
and the symmetry preserving regularization with higher covariant derivatives give different results. 

The sum of the regularized diagrams 1), 2), 3), their crossed partners and diagram  11) regularized 
by subtracting the analogous expressions with a heavy mass $\Lambda$ has the form
\begin{eqnarray}
&& \frac{e^3 m_{\phi }^2 M_Z}{M_W \sqrt{M_Z^2-M_W^2}}  
\int \frac{d^4 k }{(2 \pi)^4} 
\Biggl\{  \left(
  \frac{ g^{\mu \nu }}{[\text{\textit{k}}^2-\Lambda ^2] [\left(\text{\textit{k}}+\text{\textit{p}}_1+\text{
   \textit{p}}_2\right){}^2-\Lambda
   ^2]}-
   \frac{ g^{\mu \nu }}{[\text{\textit{k}}^2] [\left(\text{\textit{k}}+\text{\textit{p}}_1+\text{
   \textit{p}}_2\right){}^2]}\right)  \nonumber\\
   &&
 \ \  +2 \text{\textit{k}}^{\mu }
   \text{\textit{k}}^{\nu }
   \Biggl[
   \frac{1}{[\text{\textit{k}}^2]}
   \left(\frac{1}{[\left(\text{\textit{k}}+\text{\textit{p}}_1\right){}^2]}+\frac{1}{[\left(\text{\textit{k}}+\text{\textit
   {p}}_2\right){}^2]}\right) 
   \frac{1}{[\left(\text{\textit{k}}+\text{\textit{p}}_1+\text{
   \textit{p}}_2\right){}^2]}
   \nonumber\\
  && \ \ 
   -\frac{1}{[\text{\textit{k}}^2-\Lambda ^2]}
   \left(\frac{1}{[\left(\text{\textit{k}}+\text{\textit{p}}_1\right){}^2-\Lambda
   ^2]}+\frac{1}{[\left(\text{\textit{k}}+\text{\textit{p}}_2\right){}^2-\Lambda ^2]}\right)
   \frac{1}{[\left(\text{\textit{k}}+\text{\textit{p}}_1+\text{
   \textit{p}}_2\right){}^2-\Lambda
   ^2]}
   \Biggr]
      \nonumber\\
   && \ \ 
   +\text{\textit{k}}^{\nu }
   \text{\textit{p}}_2^{\mu } 
   \Biggl[
   \frac{2}{  [\text{\textit{k}}^2] [\left(\text{\textit{k}}+\text{\textit{p}}_2\right)
   {}^2] [\left(\text{\textit{k}}+\text{\textit{p}}_1+\text{
   \textit{p}}_2\right){}^2]} \nonumber\\
    && \ \ -
   \frac{2}{[\text{\textit{k}}^2-\Lambda ^2][\left(\text{\textit{k}}+\text{\textit{p}}_2\right)
   {}^2-\Lambda ^2]  [\left(\text{\textit{k}}+\text{\textit{p}}_1+\text{
   \textit{p}}_2\right){}^2-\Lambda
   ^2]}\Biggr]
   \nonumber\\
   && \ \
   +\text{\textit{k}}^{\mu }
   \text{\textit{p}}_1^{\nu } 
   \Biggl[
   \frac{2}{[\text{\textit{k}}^2] [\left(\text{\textit{k}}+\text{\textit{p}}_1\right)
   {}^2]  [\left(\text{\textit{k}}+\text{\textit{p}}_1+\text{
   \textit{p}}_2\right){}^2]}
   \nonumber\\
   && \ \ 
   - \frac{2}{[\text{\textit{k}}^2-\Lambda ^2] [\left(\text{\textit{k}}+\text{\textit{p}}_1\right)
   {}^2-\Lambda ^2]  [\left(\text{\textit{k}}+\text{\textit{p}}_1+\text{
   \textit{p}}_2\right){}^2-\Lambda ^2]}\Biggr]   \Biggr\}.
\label{d1to621}
\end{eqnarray}
Rescaling $k\to q \Lambda$, expanding the integrand in inverse powers of $\Lambda$ and 
interchanging the integration and summation we obtain in the limit $\Lambda\to\infty$:
\begin{eqnarray}
\frac{e^3  m_{\phi }^2
   M_Z  \Lambda^{D-4}g^{\mu \nu }}{D  M_W
   \sqrt{M_Z^2-M_W^2}} 
  \int \frac{d^D q }{(2 \pi)^D} \frac{
   2 (D-6) q^4-3 (D-4) q^2+D-4 }{q^4 \left(q^2-1\right)^3},
\label{resc1}
\end{eqnarray}
which is easily integrated to give exactly zero for $D=4$.

\medskip

Because of the complicated expressions below we only give the rescaled parts in the
$\Lambda\to\infty$ limit for remaining diagrams.

The rescaled expressions of diagrams 4)-9) give vanishing integrands in 
the $\Lambda\to\infty$ limit. 
The rescaled part of diagram 10) plus its crossed partner  in the limit $\Lambda\to\infty$ 
reduces to
\begin{eqnarray}
- \frac{8 (D-1) e^3 M_W M_Z \Lambda^{D-4}}{\sqrt{M_Z^2-M_W^2}}
  \int \frac{d^D q }{(2 \pi)^D}  \frac{\left(1-2 q^2\right)^2 q^{\mu } q^{\nu }}{q^6
   \left(q^2-1\right)^3}.
\label{resc2}
\end{eqnarray}
The analogous expression for diagram 12) reads
\begin{eqnarray}
\frac{2 e^3 M_W M_Z \Lambda^{D-4}}{\sqrt{M_Z^2-M_W^2}}
  \int \frac{d^D q }{(2 \pi)^D}  \frac{q^2 \left((2 D-3) q^2-D+2\right) g^{\mu \nu
   }+\left((4 D-3) q^2-1\right) q^{\mu } q^{\nu }}{q^6
   \left(q^2-1\right)^2}.
  \label{resc3}
\end{eqnarray}
The rescaled part for diagrams 13), 14) and their crossed partners  in the 
limit $\Lambda\to\infty$ sum up to
\begin{eqnarray}
 \frac{2 e^3 M_W M_Z \Lambda^{D-4}}{\sqrt{M_Z^2-M_W^2}} \int \frac{d^D q }{(2 \pi)^D}  \frac{q^2 g^{\mu \nu }-q^{\mu } q^{\nu }}{q^6
   \left(q^2-1\right)}.
  \label{resc4}
\end{eqnarray}
It is easily verified that the sum of integrals in Eqs.~(\ref{resc2})-(\ref{resc4}) 
give exactly zero for $D=4$.

Thus, the sum of all diagrams regulated by applying higher covariant derivatives in the
$\Lambda\to\infty$ limit exactly coincides with the sum of the corresponding dimensionally 
regularized diagrams obtained using the standard Feynman rules of Ref.~\cite{Aoki:1982ed}, 
taken at $D=4$. Using FeynCalc  \cite{Mertig:1990an,Shtabovenko:2016sxi} we checked that we indeed 
reproduce the old finite gauge-independent result.

\medskip

Thus we confirm once again that the problem raised in 
Refs.~\cite{Gastmans:2011ks,Gastmans:2011wh,Gastmans:2015vyh} originates from the incorrect 
treatment of the cancelling divergent integrals.
We also notice here that it is trivial to check by using FeynCalc 
\cite{Mertig:1990an,Shtabovenko:2016sxi} that vanishing results are generated if 
dimensional regularization is  applied to the expressions of Eqs.~(82) and (88) 
of Ref.~\cite{Wu:2017rxt} which are claimed in that work to be the source of the discrepancy 
between unitary and $R_\xi$ gauges if these expressions are treated more carefully.

We hope that we could convince the reader that our study refutes the reiterated claims 
of Ref.~\cite{Wu:2017rxt}  that the unitary and $R_\xi$ gauges lead to different results 
for the W-boson loop  contribution to the Higgs decay into two photons and puts 
this issue at rest, finally.

\acknowledgments
This work was supported in part  by the DFG and NSFC through funds provided to the 
Sino-German CRC 110 ``Symmetries and the Emergence of Structure in QCD'' (NSFC Grant No.  
11621131001, DFG Grant  No. TRR110),  by  the  VolkswagenStiftung (Grant No. 93562), by 
the CAS President's International Fellowship Initiative (PIFI) (Grant No. 2018DM0034) 
and by the Georgian Shota Rustaveli National Science Foundation (Grant No. FR17-354).


\end{document}